\documentclass[11pt]{article}
\usepackage{lipsum}  %
\usepackage{tabularx}
\usepackage[utf8]{inputenc}
\usepackage{hyperref}
\usepackage{setspace}
\usepackage{palatino}
\usepackage{graphicx}
\usepackage{float}
\usepackage{titling} 
\usepackage{multirow}
\usepackage{lscape}
\usepackage{amsmath}
\usepackage{amssymb}
\usepackage{subcaption}

\usepackage[a4paper, total={6in, 9.5in}]{geometry}
\fontfamily{ppl}\selectfont 

\usepackage[american]{babel}

\usepackage{setspace}

\usepackage[authoryear,round]{natbib}
\usepackage{color}

\title{Characterizing Human Actions in the Digital Platform by Temporal Context}
\author{
Akira Matsui\\
    \href{amatsui@rieb.kobe-u.ac.jp}{
        \texttt{amatsui@rieb.kobe-u.ac.jp}
        } %
\and
Emilio Ferrara\\
    \href{emiliofe@usc.edu}{
        \texttt{emiliofe@usc.edu}
        } %
}
\date{\today}

\begin{document}
{\setstretch{.8}
\maketitle

\begin{abstract}
Recent advances in digital platforms generate rich, high‑dimensional logs of human behavior, and machine learning models have helped social scientists explain knowledge accumulation, communication, and information diffusion. Such models, however, almost always treat behavior as sequences of actions, abstracting the inter‑temporal information among actions. To close this gap, we introduce a two‑scale Action‑Timing Context (ATC) framework that jointly embeds each action and its time interval. ATC obtains low‑dimensional representations of actions and characterizes them with inter- temporal information. We provide three applications of ATC to real‑world datasets and demonstrate that the method offers a unified view of human behavior. The presented qualitative findings demonstrate that explicitly modeling inter‑temporal context is essential for a comprehensive, interpretable understanding of human activity on digital platforms.
\end{abstract}

}

\section{Introduction}\label{sec:intro}

Human temporal behaviors contain a wide range of valuable information. Mining human physical, social, or economic behavior from dynamic perspectives has contributed greatly to our understanding of human mobility~\citep{song2010limits,gonzalez2008understanding,yan2017universal,schneider2013unravelling}, social phenomena~\citep{sapienza2015,pecora2017,guidict2019,holme201297,vazquez2006modeling}, and consumption behavior~\citep{platzer2016ticking,chen2020understanding}. However, most human dynamic‑behavior models focus only on the sequence of users' actions, abstracting the intervals between actions (i.e., inter‑temporal information).

Statistical time‑series models, for instance, study the variation of values in the data over time; however, such models do not explicitly capture the interdependence between actions and their intervals. While some point‑process models incorporate intervals, they use them to predict only a single or a few event types rather than to characterize diverse human actions enriched with temporal information from massive data~\citep{zhao2015seismic,mei2017neural}. Therefore, in contrast with the sophisticated advancement of statistical behavior models, understanding human behavior from the perspective of inter‑temporal context remains a difficult and often elusive goal. We perform actions in many different contexts—from using smartphones to walking across campus. Studying these situations can help us understand what human actions are like. Even the same action can differ depending on when and where it happens.

Time intervals between actions provide crucial contextual information, and much literature shows that they can reveal human cognitive states~\citep{stanovich2000individual,graesser1997discourse,wickens2021information,kahneman2011thinking}. The most famous theory in the literature and popular science is ``Dual‑process theory'' introduced by~\citet{graesser1997discourse} and developed by~\citet{kahneman2011thinking}. In addition to the cognitive processes that generate action intervals, the statistical properties of interval distributions have attracted attention from researchers, especially in network science~\citep{holme201297,masuda2016guide,vazquez2006modeling,barabasi2005origin,oliveira2005darwin}. This line of literature suggests that studying the temporal context of actions in terms of intervals fosters our understanding of temporal human behavior.
This calls for a model that integrates time‑interval information into actions to account for context when mining and interpreting actions. However, modeling user actions and their intervals holistically is challenging because actions are categorical data, whereas intervals are numerical. One natural solution is to discretize intervals using time bins~\citep{wang2016unsupervised}, but defining and validating those bins objectively is cumbersome. In addition, even if we successfully discretize intervals, incorporating interval information into action analysis is not trivial.

In summary, existing models often omit crucial information about human behavior within the time intervals between actions, resulting in a limited, objective understanding of the temporal structure of human behavior. To overcome these challenges, we propose a framework that embeds user actions alongside their inter-temporal intervals. The methods proposed in this paper utilize statistical mixture models for objective interval discretization and word embedding techniques to generate unified, low-dimensional representations that capture temporal information. The embeddings obtained by our method allow us to extract and contextualize human behavior from large datasets.

The presented paper first leverages the statistical properties of intervals that have been well explored. Our model estimates the distribution as a mixture of exponential distributions and uses the estimated parameters to discretize intervals. This statistical procedure ensures objectivity, avoids manual hyperparameter selection for bins, and guarantees that the discretized intervals capture the nature of the observed data. We then construct action sequences in which n-grams represent the interdependence between actions and intervals. Our framework calculates a low‑dimensional representation of user actions with interval information using a word‑embedding algorithm. Lastly, we study the inter‑temporal context for each action using the obtained embedding vectors. We introduce an interpretable measure, the action timing context (ATC), which identifies whether a given action belongs to a long‑term (slow‑paced) context or a short‑term (fast‑paced) context.  The novelty of our work lies in embedding time intervals alongside actions and characterizing each action with its associated time interval. Our paper serves to introduce and establish this baseline model for temporal behavior representation.

\begin{figure}[ht]\centering
    \includegraphics[width=.9\columnwidth]{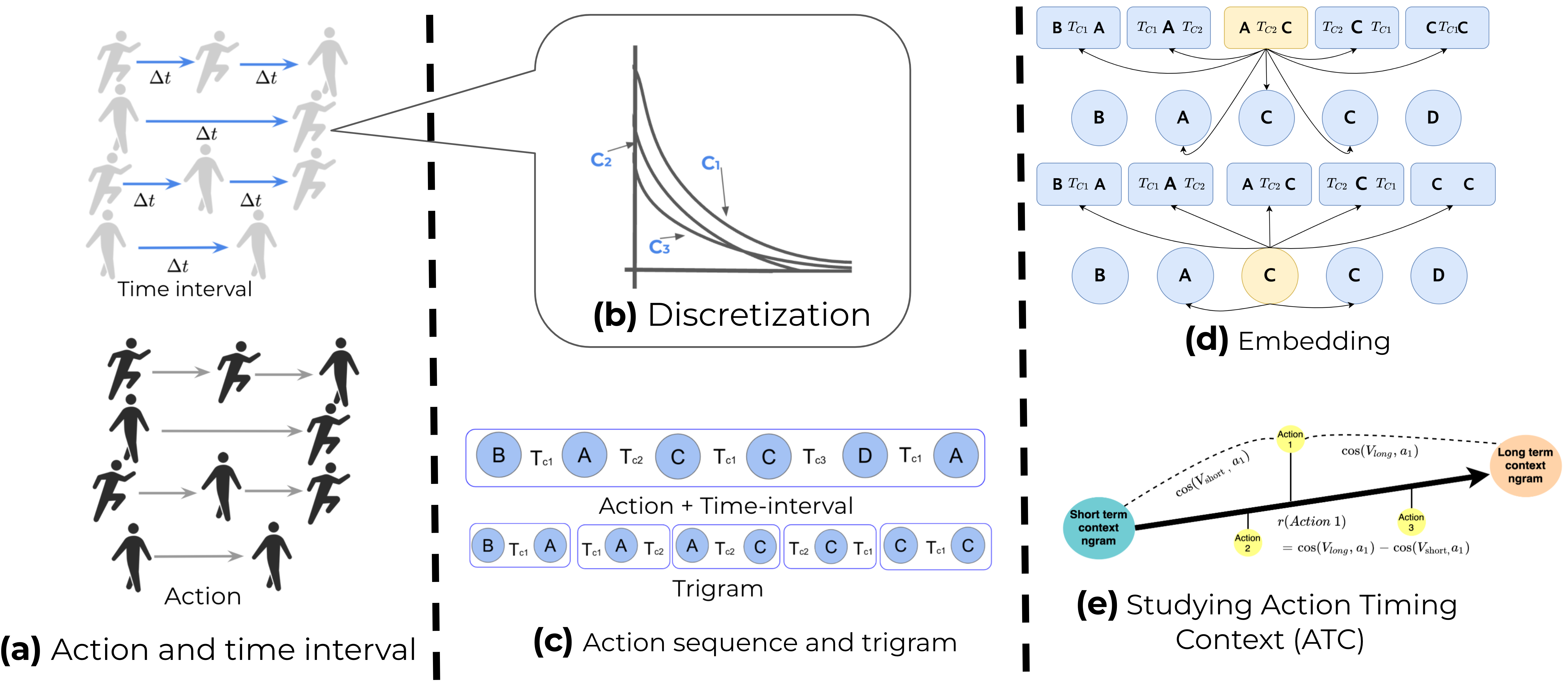}
    \caption{Schematic of studying the inter-temporal context of user actions}
\label{fig:schematic}\par
\medskip
\begin{minipage}{.95\columnwidth} %
{\footnotesize Note: \textbf{(a)} The action sequence of each user and calculate time intervals between two consecutive actions.\textbf{(b)} The mixture distribution of time intervals to discretize the time intervals. \textbf{(c)} The sequences of action $+$ time interval and construct trigrams from those sequences.
\textbf{(d)} The embedding vector of actions and trigrams learned from the n-gram sequence. \textbf{(e)} The inter-temporal context of action extracting Action Timing Context (ATC) from obtained embedding vectors.}
\end{minipage}
\end{figure}

For clarity, we illustrate the schematic of our study in Figure~\ref{fig:schematic}. Throughout this paper, we use the term ``action'' to refer to a discrete interaction event recorded on a digital platform (e.g., launching an app), not human‑pose action recognition. In addition, we use ``action'' to refer to a single unit of observed behavior and ``time interval'' to refer to the gap between two consecutive actions. The remainder of this paper is structured as follows. In Section~\ref{sec:related}, we first review the background of the literature that motivates our study. Then we detail the proposed framework (ATC) for embedding actions with their inter‑temporal context in Section~\ref{sec:method}. After Section~\ref{sec:data_exp_sett} introduces the datasets and experiment, Section~\ref{sec:empirical} presents the three application examples of the proposed method. Finally, Section~\ref{sec:conclusion} concludes and outlines directions for future research.

\section{Related Research and Theoretical Background}\label{sec:related}

This section reviews the related research to our study, spanning behavior models with inter‑temporal information and machine‑learning models that extract the essence of human behavior from complex, high‑volume data. We start this section by discussing the literature on human behavior modeling that utilizes time intervals between human actions. This line of literature has pointed out that studying time intervals can reveal regularities common to a wide variety of human behavioral phenomena and help us understand the cognitive states of humans who execute their actions. We also review the literature on user behavior modeling that leverages word‑embedding models. 

\subsection{Inter‑temporal information, actions, and human cognitive states}

The inter‑temporal aspect of human behavior contains ubiquitous regularities underlying human behavior~\citep{barabasi2005origin}. Since the time intervals of consecutive actions are definable for any type of human action, understanding the statistical properties of them has been an important issue in revealing the regularities behind human behavior. Literature across a wide variety of research areas is interested in this realm and finds that mining the inter‑temporal information of actions is reflected in human mobility patterns~\citep{song2010limits,gonzalez2008understanding,yan2017universal,schneider2013unravelling}, social phenomena~\citep{sapienza2015,pecora2017,guidict2019,holme201297,vazquez2006modeling}, and consumption behavior~\citep{platzer2016ticking,chen2020understanding}. Consequently, the study of time‑interval distributions has attracted the attention of many researchers~\citep{holme201297,masuda2016guide,vazquez2006modeling,barabasi2005origin,oliveira2005darwin}, and the literature has proposed two classes of models: queue models~\citep{barabasi2005origin} and modulated Markov processes~\citep{holme201297,masuda2016guide}.  

In addition, psychology literature points out that studying interactions between human actions and their time intervals reveals deep mechanisms of human cognitive states~\citep{stanovich2000individual,graesser1997discourse,wickens2021information,kahneman2011thinking}. Dual process theory belongs to this line of literature and dichotomizes human cognitive states into System 1 and System 2~\citep{graesser1997discourse,stanovich2000individual}. This theory models the mode of cognitive state with regard to decision‑making processes and argues that System 1 reaches decisions quickly, while System 2 takes longer. We employ this theory to characterize actions based on the cognitive state in which they are executed, embedding the time interval between those actions.  

We build our methodological and conceptual backbone on the above diverse literature and propose the framework that extracts and characterizes human actions based on the cognitive states of the humans who execute those actions, using massive, complex data. Section~\ref{sec:method} proposes the ATC framework to study the temporal structure of users' actions on this simple yet solid foundation.  

\subsection{User modeling with word embedding}

This study models human behavior with temporal information represented by the time intervals between actions. To this goal, we use word‑embedding models, which are neural network models first proposed in natural language processing research. Word‑embedding models are a class of machine learning models that extract low‑dimensional representations from documents. Word embeddings have been mainly used to obtain low‑dimensional representations of word semantics~\citep{Mikolov2013Efficient,mikolov2013dist}. They are also known to capture biases hidden in text data~\citep{gonen2019lipstick,bolukbasi2016man,garg2018word,caliskan2017semantics} and semantic dynamics~\citep{hamilton2016diachronic,di2019training}.  

The applications of such embedding techniques extend beyond natural‑language processing to user modeling and there is an emerging trend in the application of word embedding to social science research \citep{matsui2022word}. Although originally designed for text analysis, this powerful tool has been applied to processes using non‑textual data, especially in social science~\citep{matsui2022word}. For example, \citet{Aceves2024} investigates the speed of human communication in terms of information density, \citet{Lewis2023} examines the semantic spaces of human languages, and \citet{Charlesworth2022} investigates historical changes in societal representations of social groups.

Recently, it has been found that embedding models can capture political orientation, which is not explicitly coded in the content of platforms. \citet{waller2021quantifying} propose a method and application to study social phenomena in the online community (Reddit) in a data‑driven way. They characterize the online community (subreddit) by users' posts and study political polarization dynamics. For modeling dynamic behavior, \citet{han2020modelling} propose an embedding model that obtains the vector representation of user dynamics from user action sequences. Such a model can also predict students' academic performance by embedding their daily behavior sequences~\citep{LiuJointly2020}.   

The major difference between our study and the studies discussed above is that our study explicitly embeds temporal structure into actions. Existing studies use action sequences to represent the dynamics of user behavior and abstract the time intervals between actions, even though, as mentioned in the introduction, temporal information is necessary to understand human behavior. Certainly, using point processes and other methods can improve prediction accuracy~\citep{du2016recurrent,Karishma2021Identifying}, but such models complicate the understanding of human behavior. In this study, we explicitly model temporal structure in the embedding of behavior by estimating the time intervals. This makes the model structure very simple and enables a unified analysis of user behavior using indicators such as ATC, as proposed in this study.

\section{Method}\label{sec:method}
This paper aims to study human behavior considering inter‑temporal information. To this end, our method seeks to obtain a low‑dimensional representation of the action sequence. We construct action sequences for each user with the actions and their time intervals, and then apply a word‑embedding method to the constructed sequences.

\subsection{Capturing inter‑temporal information with time bins}

This subsection discusses how we estimate the time intervals and convert them into the action sequence using time bins. While calculating time intervals between consecutive actions is straightforward, it is not trivial to discretize this continuous‑time value for inclusion in the action sequence. A natural solution to this discretization problem is to treat some time bins as hyperparameters. For example, \citet{wang2016unsupervised} use predefined bins such as ($[0\text{–}1\text{min}], [1\text{min–}10\text{min}], [10\text{min–}60\text{min}], [60\text{min}<]$) and then classify calculated time intervals into those bins. However, it is not always easy to create an appropriate set of bins without prior information. Without considering the characteristics of the time intervals in the data, one might miss the inter‑temporal information between actions.

To overcome this point, we estimate the statistical properties of the time intervals in the data to construct time bins based on the estimated parameters, rather than using bins prepared from a given hyperparameter. Recent developments in the literature reveal that a mixture of exponential distributions with a few components can fit the time intervals of various empirical data well~\citep{okada2020long}. The goal of this time‑bin construction is to capture the users’ behavioral states from their time intervals and represent these states with the time bins. To this end, we study the behavioral states by estimating mixture distributions.    

\subsection{Time‑interval‑bin estimation by a mixture of exponential distributions}\label{sec:time_bins_est}

We estimate the distribution of time intervals between actions and construct time bins based on the estimated distribution, assuming that the observed time intervals follow a mixture of exponential distributions. For estimation, we utilize the Expectation-Maximization (EM) algorithm and a model‑selection criterion.

To study the mixture of exponential distributions from the observed time intervals between actions $x=\{x_{1}, \ldots, x_{n}\}$, we employ the EM algorithm to estimate the parameters of the interval distributions, modeled as a mixture with $K$ components.

\begin{equation}\label{eq:overall_dist}
\mathbf{f}(x)=\sum_{k=1}^{K}\pi_{k}f_{k}(x;\lambda_{k}),
\end{equation}

where $\pi_{k}\ge 0$, $\sum_{k=1}^{K}\pi_{k}=1$, $f_{k}$ denotes the probability‑density function of an exponential distribution, and $\lambda_{k}$ is the rate parameter of component $k$. We select $K$ based on the decomposed normalized maximum‑likelihood (DNML) codelength, following \citet{okada2020long}.

We assign the time intervals between actions to the time bins based on the estimated mixture. We construct as many time bins as there are components in the mixture distribution, plus an additional bin for the zero time interval, $T_0$. In other words, when we have $K$ components, we construct $K+1$ bins, $\{T_0, \ldots, T_{K}\}$. For a given time interval $x_i > 0$, we find the time bin $T^{*}$ for $x_i$ using the results of the estimation as 
\begin{align}\label{eq:time_bin}
    T^{*} = \operatorname*{argmax}_{1\le l\le K} \frac{\pi_l f(x_{i} ; \lambda_l)}{\sum_{k=1}^{K} \pi_{k} f(x_{i} ; \lambda_{k})}.
\end{align}
Since the time‑bin attribution cannot be calculated for $x_i=0$; therefore, we create a special time bin, $T_0$, for $x_i=0$.

\subsection{Constructing action n‑gram sequences}

To represent the context of users’ actions, we construct sequences of actions and their time intervals in chronological order and then create n‑grams from these sequences.

\subsubsection{Action sequence}
With the actions and time intervals between them, we construct action sequences for each user. Let user $i$ have the actions $\{A_{i,j}\}_{j=0}^{J}$. Their $j$‑th action $A_{i,j} \in \{A_{1} \ldots A_{M}\}$, where $M$ is the number of unique actions, has a time interval $x_{ij}$, the interval between $A_{i,j}$ and $A_{i, j+1}$ (for $j=0, \ldots, J-1$). Using the constructed time bins, we assign $x_{ij}$ to a bin $T_{ij}$ (for $j=0, \ldots, J-1$), which is one of $\{T_0, \ldots, T_{K}\}$. With user $i$’s actions $\{A_{i,j}\}_{j=0}^{J}$ and their time bins $\{T_{i,j}\}_{j=0}^{J-1}$, we construct the action sequence as
$$A_{i,0}T_{i,0}A_{i,1}T_{i,1}A_{i,2}T_{i,2} \ldots A_{i,J-1}T_{i,J-1}A_{i,J}.$$
We construct such an action sequence for each user and then create n‑grams from these sequences.

\subsubsection{Action n‑gram}\label{sec:action_ngram}

Our action n‑gram captures the context of users’ actions, representing the order of actions and their time intervals. For an action sequence\footnote{For simplicity, we use $A_{j}$ and $T_{j}$ for $A_{i,j}$ and $T_{i,j}$, respectively, omitting the user index $i$.}
{\small $A_{1}T_{1} \ldots T_{J-1}A_{J}$},
we create n‑grams by separating $A_{j}$ and $T_{j}$. In this study, we use trigrams ($n=3$) of the action sequence; they are
$$
{\{
A_{1}T_{1}A_{2},\ 
T_{1}A_{2}T_{2},\ 
\ldots
T_{J-2}A_{J-1}T_{J-1},\ 
A_{J-1}T_{J-1}A_{J}
\}.}
$$
While others may prefer to treat the pair of an action and its time interval as a single unit (e.g., $A_{J}T_{J}$), creating n‑grams that separate the two is advantageous for studying the time‑interval context of the action (see Section~\ref{sec:t_int_context}). 

\subsection{Embedding action n‑gram sequences}

We use the Skip‑gram with negative sampling (SGNS) algorithm to obtain low‑dimensional representations of actions and n‑grams~\citep{Mikolov2013Efficient,mikolov2013dist}. This section discusses the prediction problem that SGNS solves in this paper to clarify the embedding vectors obtained from the action sequences. The SGNS models the distribution $p(d \mid w, c)$, where $d$ equals 1 when a pair consisting of word $w$ and context $c$ is observed in the data, and 0 otherwise. We aim to maximize the log‑likelihood function $\mathcal{L}$ defined as
\begin{align}
  \mathcal{L} = \sum_{(w, c) \in \mathcal{D}} \log p(d=1\mid w, c) + k\, \mathbb{E}_{\bar{w} \sim q} \bigl[\log p(d=0 \mid \bar{w}, c)\bigr],
\end{align}
where $q$ is the noise distribution in negative sampling, and $k$ is the sample size from the noise.\footnote{This log‑likelihood maximization with the noise distribution is considered Noise Contrastive Estimation (NCE)~\citep{gutmann2010noise}; SGNS is a variation of NCE.} The SGNS calculates $p(d=1 \mid w, c)$ with the sigmoid function $\sigma(v_{c} \cdot v_{w})$, where $v_{w}, v_{c} \in \mathbb{R}^{D}$. In other words, SGNS seeks the parameters $v_{w}$ and $v_{c}$ that maximize this probability; these vectors are the embeddings of interest. 

\subsubsection{``Word’’ and ``Context’’ in this paper}
This paper treats actions and n‑grams of action sequences as either words or contexts. Note that we treat actions as both words and contexts, and we do the same for n‑grams. We calculate the conditional probability of actions or n‑grams given actions or n‑grams to obtain low‑dimensional representations. For implementation, we use ngram2vec~\citep{zhao2017ngram2vec}, a modified version of word2vecf~\citep{levy2014dependency}.

\subsection{Extracting action‑timing context (ATC) using n‑gram actions}\label{sec:t_int_context}

To contextualize actions with inter‑temporal information, we use the embedding vectors of the n‑grams. As discussed in Section~\ref{sec:action_ngram}, n‑grams contain elements representing an action between time intervals, such as $TAT$. We leverage the embedding vectors of this type of n‑gram as references representing the inter‑temporal context. 

\subsubsection{Constructing the reference vectors}
First, we create two types of reference n‑grams: long‑ and short‑time‑interval contexts. Let $T_{long}$ and $T_{short}$ denote the longest and shortest time‑interval bins, respectively (defined in Section~\ref{sec:time_bins_est}). We calculate the reference vectors for the long‑ and short‑interval contexts as
\begin{align}\label{eq:l-s vector1}
    v_{long} \equiv \frac{1}{|\mathcal{V}_{long}|} \sum_{v \in \mathcal{V}_{long}} v,
\end{align}
where $\mathcal{V}_{long}$ is the set of embeddings of trigrams in which actions occur between the longest time intervals,
\begin{align}\label{eq:l-s vector2}
     \mathcal{V}_{long} \equiv \{
     T_{long}AT_{long}
     \}_{A \in \mathcal{A}},
\end{align}
and $\mathcal{A}$ is the set of all actions. We construct $\mathcal{V}_{short}$ analogously using $T_{short}$.

\subsubsection{Definition of ``long‑term context’’ and ``short‑term context’’}
The reference vectors $v_{long}$ and $v_{short}$ represent actions taken between long‑term or short‑term intervals, respectively. For example, when an action $A$ is similar to $v_{long}$, it means that the user takes a long break (interval) before or after that action. We interpret such an action as one executed after a long waiting period. Conversely, when an action $A$ is similar to $v_{short}$, it means that users tend to execute it after a short waiting period.

\subsubsection{Aligning actions into long‑ vs. short‑term context}
\begin{figure}[!ht]\centering
    \includegraphics[width=0.7\columnwidth, height=0.28\columnwidth]{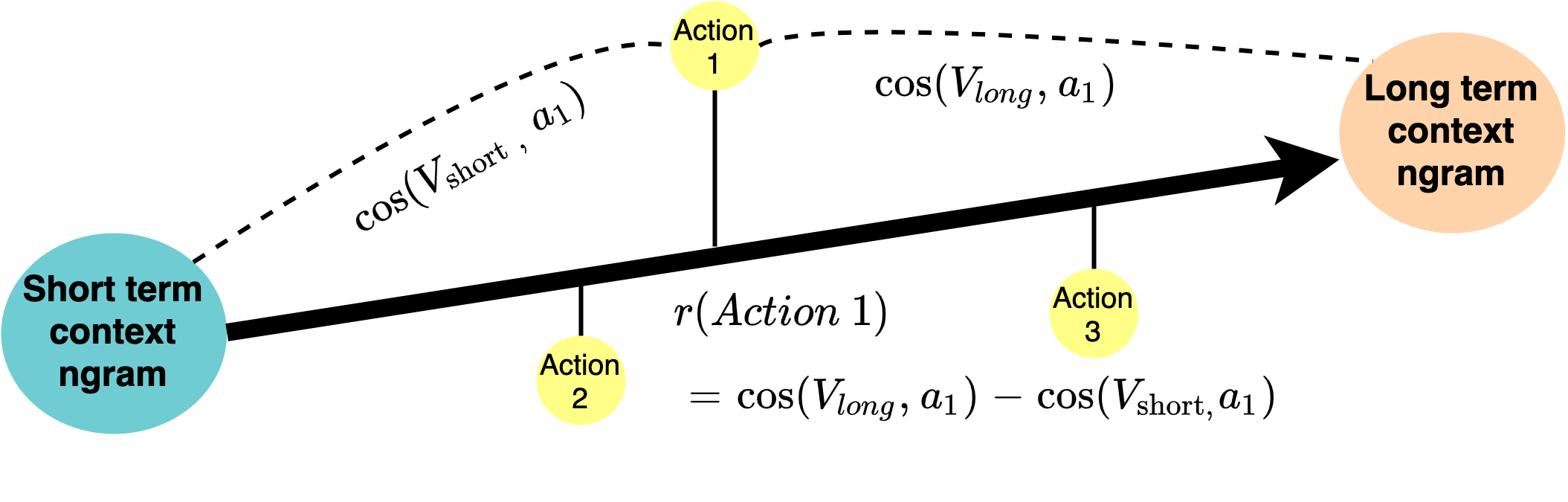}
\caption{Extracting action‑timing contexts with n‑gram actions}
\label{fig:relative_dif}\par
\medskip
\begin{minipage}{\columnwidth}
{\small\it Note:}
We calculate the relative distance from the long‑term and short‑term context n‑grams. First, we construct the vectors for the long‑term ($v_{long}$) and short‑term ($v_{short}$) context n‑grams (Equations~\ref{eq:l-s vector1} and~\ref{eq:l-s vector2}). Then, for each action vector $a$ of $Action\ 1$, we calculate the difference between the cosine similarities $\cos(v_{long}, a)$ and $\cos(v_{short}, a)$. This relative distance $r(Action\ 1)$, defined in Equation~\ref{eq:relative_distance}, indicates the context to which the action belongs. A large $r(Action\ 1)$ suggests that the action is used in a long‑term context.
\end{minipage}
\end{figure}

To study the time‑interval context of a given action $Action\ i$, we calculate the relative distance $r(A)$ between the two reference vectors for each action of interest,
\begin{align}\label{eq:relative_distance}
 r(Action\ i) = \cos(v_{long}, a) - \cos(v_{short}, a),
\end{align}
where $a$ is the embedding vector of action $A$, and $\cos$ denotes the cosine similarity. When $r(Action\ i)$ is large, $Action\ i$ is executed in the long‑term context; otherwise, $Action\ i$ is in the short‑term context. We provide a schematic illustration of this relative distance in Figure~\ref{fig:relative_dif} for the case where $i=1$.

\section{Data \& Experiment Settings}\label{sec:data_exp_sett}

This section describes the dataset and experiment settings for the embedding model. We employ three datasets to investigate a wide range of human behavior in terms of inter-temporal context using the ATC discussed in Section~\ref{sec:t_int_context}. 

\begin{table*}[!htbp]
  \centering
  \small
  \caption{Basic dataset statistics}
  \label{table:basic_statistic} 
  \begin{tabularx}{.95\columnwidth}{lXXX}
    \textbf{Dataset Statistics} & \textbf{App Usage} & \textbf{MoocData} & \textbf{StudentLife} \\
    \hline
    Observation period & 1 week & 2 years$+$11 days & 11 weeks\\
    Observation date & Jun'16 & Jun'15--Jun'17 & Mar'13--May'13\\
    Observation field & Smartphone & MOOC platform & University campus\\
    Environment & Digital device & Digital platform & Digital device and real place\\
    \# unique actions & $1,696$ & $22$ & $800$\\
    \# total users & $871$ & $225,642$ & $49$\\
    \# total actions & $4,171,950$ & $42,110,402$ & $219,360$\\
    Avg.\ \# actions per user & $4,789.83$ & $186.62$ & $4,476.73$\\
    Avg.\ \# unique actions per user & $1.94$ & $9.74$ & $16.326$\\
    \# components of mixed dist & $3$ & $3$ & $3$\\
    \hline
  \end{tabularx}

  \begin{minipage}{.95\columnwidth}
    \footnotesize \textit{Note}: Basic statistics of the three datasets used for our analysis after the preprocessing described in Sec.~\ref{sec:data_exp_sett}. ``\# components'' is determined by DNML, as discussed in Sec.~\ref{sec:time_bins_est}.
  \end{minipage}
\end{table*}

\subsection{Datasets}

We apply our method to three datasets. First, we use the app usage history dataset by~\citet{feng2019dropout} to study the correspondences between inter-temporal context differences (ATC) and app categories. Then, we use ATC to study behavioral differences between individuals, demonstrating that our method captures human behavior in terms of inter-temporal context. For this analysis, we use clickstream data from a Massive Open Online Course (MOOC) platform by~\citet{yu2018smartphone}. Lastly, we examine ATC dynamics using the StudentLife sensing data by~\citet{wang2014studentlife}. 

\subsubsection{App Usage Dataset}
We use the smartphone app usage history data~\citep{feng2019dropout}. Smartphone apps are essential tools in modern daily life, providing functions such as games and health tracking. These apps likely have different ATCs depending on their purpose. The dataset provides users' smartphone app usage histories, including timestamps and app categories. We treat each app usage as an action and label it with its category. 

\subsubsection{MoocData}
MoocData comprises users' clickstream data set on a Mooc platform published by ~\cite{yu2018smartphone}. The dataset contains logs of users’ learning activities, including which platform functions they use. It also serves as a dataset for dropout prediction. We treat each clickstream event as an action and compute time intervals based on timestamps. Also, we utilize this subset to compare behavioral differences between two types of users: dropout vs.\ non‑dropout.

\subsubsection{StudentLife}
The StudentLife dataset provides a wide range of behavioral data collected automatically via participants’ smartphones. \citet{wang2014studentlife} recruited students and tracked their behavior on campus during one academic term. The dataset includes smartphone usage, eating behavior (e.g., lunch), and physical activities (e.g., walking).
We first align these actions based on timestamps and compute the time intervals between them. This dataset also provides Ecological Momentary Assessment (EMA) survey data, reporting students’ physiological states throughout the period. We use three questionnaires to study correlations between ATC and students’ psychological states.

\begin{table}[!hbp]
  \small
  \caption{EMA Questions}
  \label{table:ema_question} 
  \centering
  \begin{tabularx}{\columnwidth}{XX}
    \textbf{Question} & \textbf{Option (scale)}\\
    \hline
    how happy do you feel? & 1.\ a little bit, 2.\ somewhat, 3.\ very much, 4.\ extremely\\
    how sad do you feel? & 1.\ a little bit, 2.\ somewhat, 3.\ very much, 4.\ extremely\\
    How are you right now? & 1.\ happy, 2.\ stressed, 3.\ tired\\
    How many hours did you sleep last night? & 18 scales (0.5‑hour grid from less than 3 hours to more than 12 hours)\\
    \hline
  \end{tabularx}
  \begin{minipage}{.99\columnwidth}
    \footnotesize \textit{Note}: EMA questions used in this paper. Participants who answer ``yes'' to ``Do you feel AT ALL happy (sad) right now?'' then answer ``How happy (sad) do you feel?'' 
  \end{minipage}
\end{table}

We utilize these datasets to demonstrate that ATC captures human behavior from an inter-temporal perspective.

\subsection{Experiment Settings}

As discussed in Sec.~\ref{sec:time_bins_est}, we estimate a mixture of time intervals between actions to construct time bins, using the implementation by~\citet{okada2020long}. For each dataset, we randomly sample 10\,k time intervals and estimate their distributions.
For embedding, we use ngram2vec~\citep{zhao2017ngram2vec} based on word2vecf~\citep{levy2014dependency}. We use SGNS to learn 300‑dimensional embedding vectors, following the settings in the implementation by \cite{zhao2017ngram2vec}. We adopt a flexible window size—two windows for bi‑grams and one window for actions, as illustrated in Fig.~\ref{fig:schematic}(d)—to ensure that our model captures dependencies among actions and time intervals, but we remove users with fewer than 10 actions.

\section{Empirical Analysis with Action Timing Context (ATC)}\label{sec:empirical}

This section reports an empirical analysis of action timing contexts using real‑world datasets. We demonstrate that action timing context extracts informative insights about human behavior from action sequences. The analysis uses three datasets: student behavior observed in a field study, smartphone app usage data, and student behavior in a MOOC. Using these datasets, we compare action timing contexts across different entity types. First, we show that the ATC of app usage depends on its category. Since applications have different purposes, these differences are reflected in their ATC. Next, we use our method to identify behavioral differences between successful and unsuccessful students, comparing ATCs of dropout and non‑dropout students on the MOOC platform. Lastly, we study transitions in ATC of student behavior in an academic environment using datasets covering physical and digital behavior over an academic term.

\subsection{ATC Differences among Smartphone App Usage}

\begin{figure}[!ht]\centering
    \includegraphics[height=.7\columnwidth]{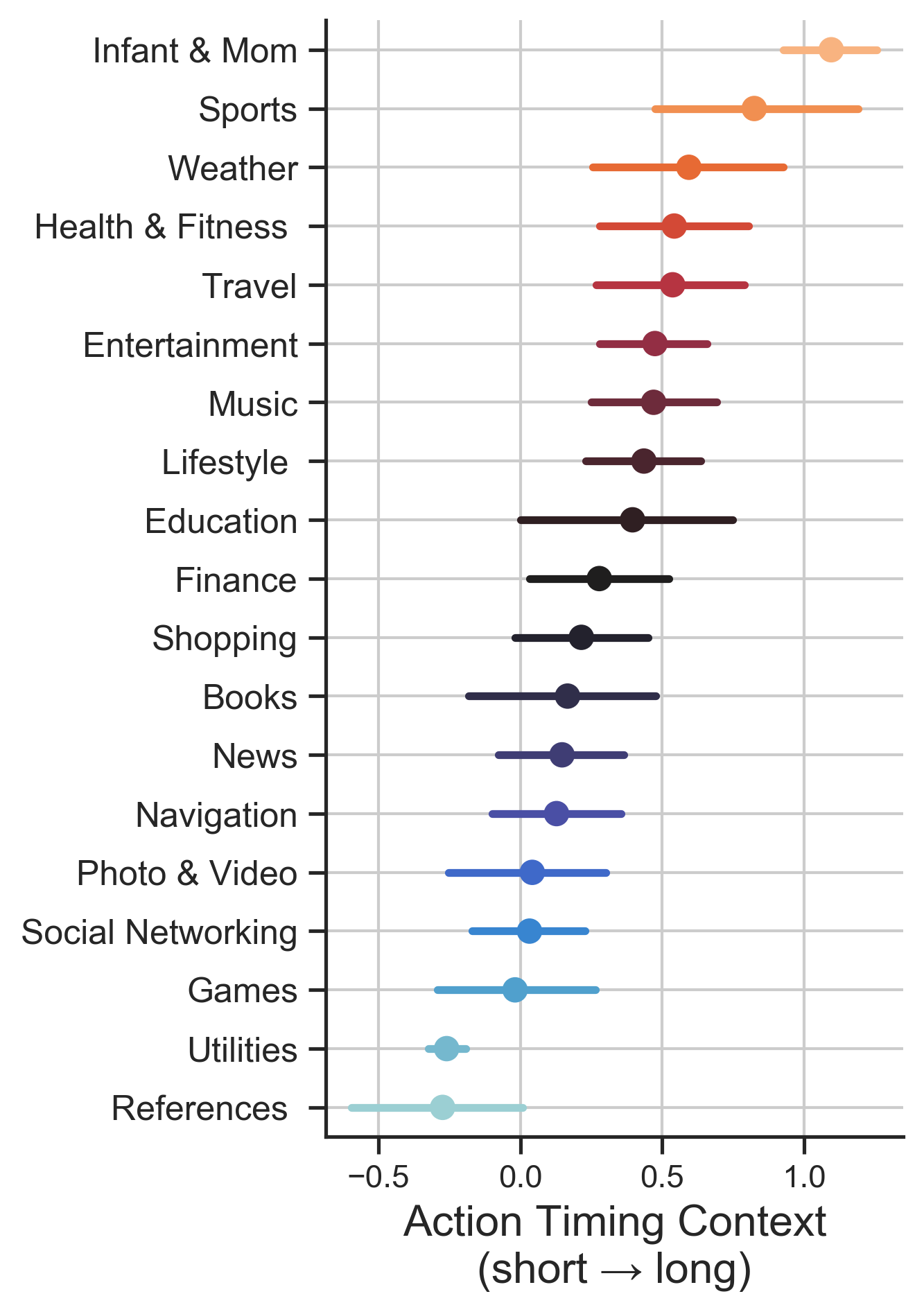}
\caption{Smartphone Apps usage context}
\label{fig:app_atc_category}\par
\begin{minipage}{.95\columnwidth} %
{\footnotesize {\it Note: } Comparisons of smartphone app usage by the action timing context (standardized on all samples). The figure plots the mean value of each category's action timing context (with 95\% confidence interval by bootstrap). While the apps for ``Infant \& Mom'' are used in the long action timing context, ``Reference'' category apps are used in the short-term context. Also, the ``Finance'' category apps are neutral in terms of the action timing, suggesting that the users use those apps in both contexts.}
\end{minipage}
\end{figure}

We first apply our method to the \textbf{App Usage Dataset} to examine whether smartphone apps differ in ATC depending on their purposes. Therefore, temporal context varies with app purpose. To study this, we calculate the mean ATC for each app category. Figure~\ref{fig:app_atc_category} shows that different categories have different mean ATCs. In the figure, the ``Reference'' category has the smallest ATC (–0.27), indicating that apps in this category, such as dictionaries, are used for short periods. By contrast, ``Infant \& Mom'' has the largest ATC (1.1), suggesting long‑term usage. Our analysis also reveals apps that do not belong to a definitive context. For example, the ATC value of finance apps is 0.27, indicating usage in both long‑ and short‑term contexts. 

\subsection{ATC Differences between Dropout and Non‑Dropout Students}
The second application of our method is the detection of behavioral differences among users based on their performance on a digital platform. We use \textbf{MoocData} to examine whether ATC can capture differences in student learning behavior on the MOOC platform and the proposed method reveals behavioral differences in an interpretable manner. We use student clickstream data from a MOOC platform, which labels each student as dropout or non‑dropout. We split the dataset accordingly, learn embeddings for each subset, and calculate ATCs, enabling comparison of the same action across student types.

\begin{figure}[!ht]\centering
    \includegraphics[height=.7\columnwidth]{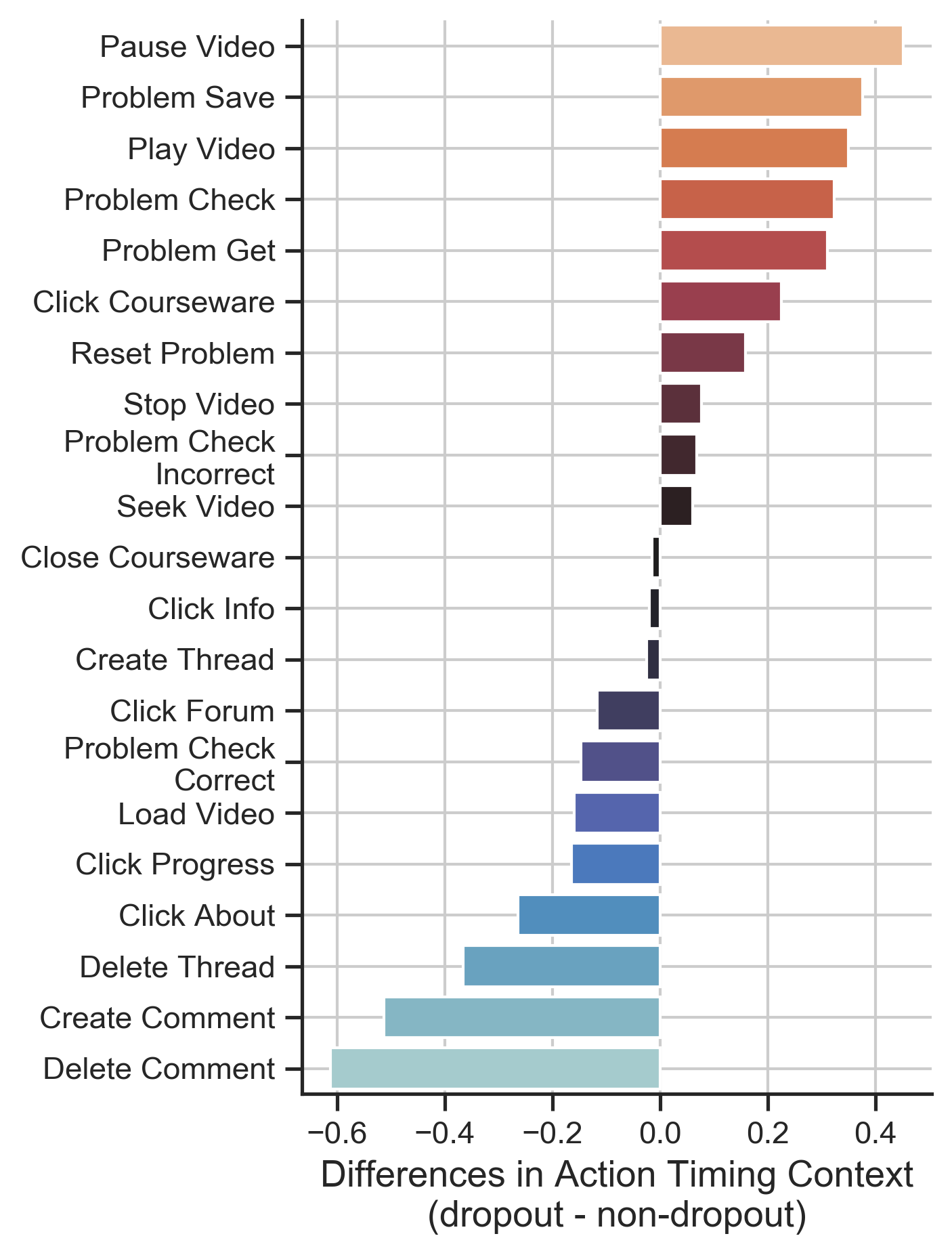}
\caption{Action timing context differences: Drop-out students VS Non-dropout students}
\label{fig:mooc_atc_diff}\par
\begin{minipage}{\columnwidth} %
{\footnotesize {\it Note: } Action timing context difference between the different types of users (standardized on each student type). We calculate the action timing context of the students in the MOOC platform. We calculate the action timing context for the students who dropped out of their course and the student. A positive difference of action represents the dropout users take that action in the long-term context, but the non-dropout students use that action in the short-term context. For example, the dropout students use ``Pause Video'' in the long-term context, but the non-dropout students use it in the short-term (difference 0.45). This difference implies that the dropout students do not often pause the course video, but the non-dropout students do so.}
\end{minipage}
\end{figure}

Figure~\ref{fig:mooc_atc_diff} plots ATC differences for identical actions between dropout and non‑dropout students, computed by subtracting dropout ATCs from non‑dropout ATCs. A positive difference implies that dropout students use the action in a long‑term context, whereas non‑dropout students use it in a short‑term context. The most evident distinction is for ``Pause Video'' (0.45): dropout students seldom pause course videos, whereas non‑dropout students pause frequently in short intervals. Conversely, negative differences in commenting (``Delete Comment'': –0.61; ``Create Comment'': –0.51) indicate that non‑dropout students spend more time commenting than dropout students. Although ATC differs across many actions, both groups are similar for ``Close Courseware'' (–0.01). Combined, these patterns highlight key behavioral differences: compared with non‑dropout students, dropout students are less likely to pause videos and spend less time commenting. 

\subsection{Capturing Behavior Dynamics by ATC}

Lastly, we use our method to study behavior dynamics and present the appliction to \textbf{StudentLife} dataset. Using this dataset, we analyze how ATC transitions over an 11‑week academic term. The dataset includes smartphone‑based traces such as physical activity, eating behavior, and app usage. We construct weekly action sequences, learn embeddings, and compute weekly ATCs.

\begin{figure}[ht]
\centering
     \subfloat[][]{
     \includegraphics[width=.5\linewidth, height=.34\linewidth]{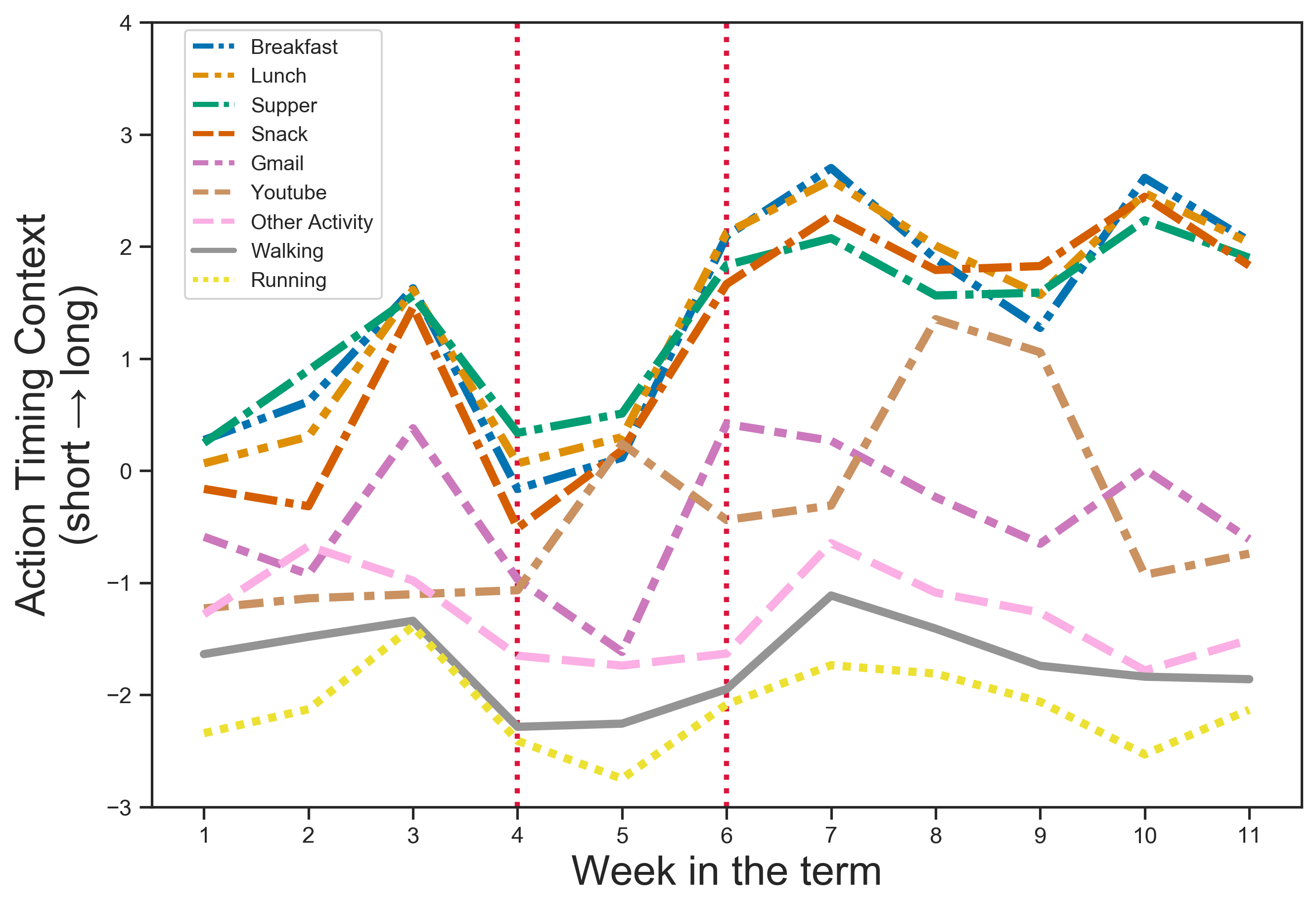}
     \label{fig:studentlife_weekly}}
     \subfloat[][]{\includegraphics[width=.38\linewidth, height=.31\linewidth]{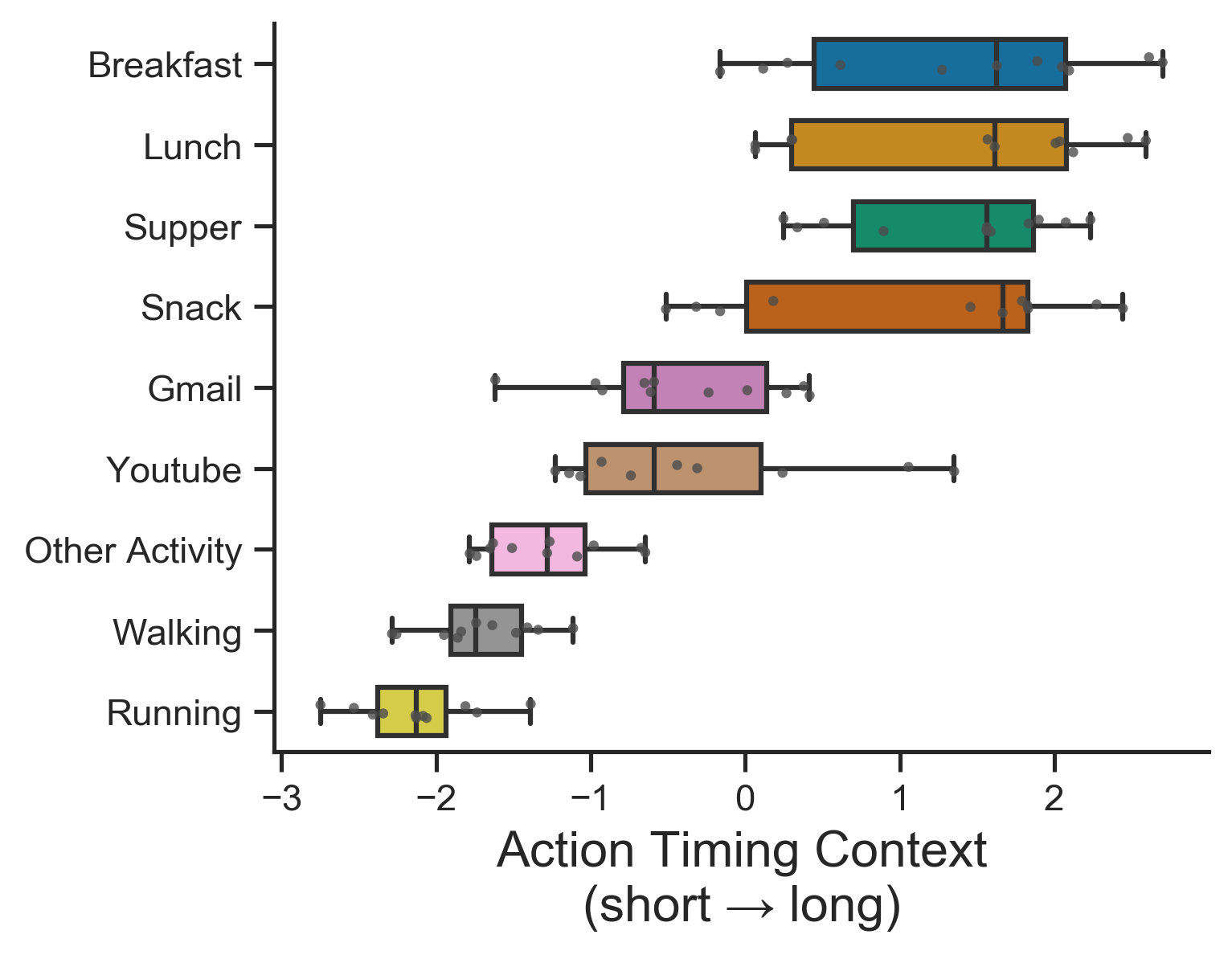}
     \label{fig:studentlife_box}}
     \caption{Transition of the action timing context of the students over the academic term}
     \label{steady_state}
    \begin{minipage}{.95\columnwidth}%
    {\footnotesize {\it Note}: Calculating the action timing context of the student behavior over the course of an academic term (11 weeks, standardized on each week). We calculate the action timing context of the actions in each week: the physical behavior (e.g., walking), the smartphone app usage (e.g., YouTube), and the eating activity (e.g., lunch). \textbf{Figure~\ref{fig:studentlife_weekly}} shows the ATCs changes over the academic terms. The area between the dot lines represents the mid-term weeks (Week4-6). While the context of physical behavior is stable, eating activity transitions from the long-term context to the short-term context. This transit implies that the important academic event off the students stride in their eating habit. \textbf{Figure~\ref{fig:studentlife_box}} plots the box plot of the action timing context of the actions, and shows that while the physical activities are stable, the eating activities are volatile (having longer bars).}
    \end{minipage}%
\end{figure}

\begin{figure}[ht]
     \centering
     \includegraphics[width=.9\linewidth]{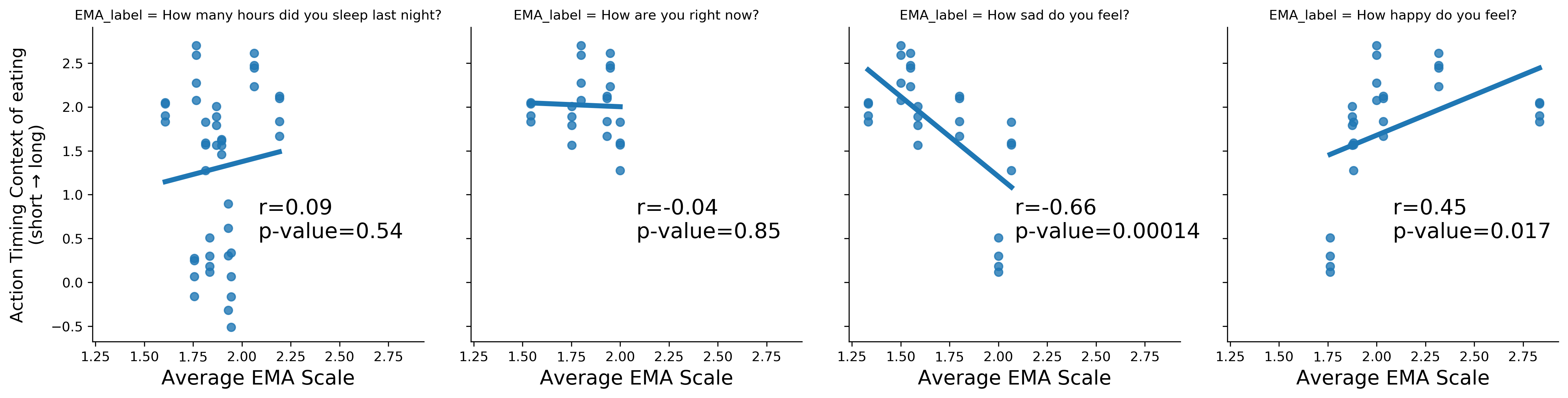}
     \caption{Correlations between ATCs of the eating actions and Physiological state}\label{fig:studentlife_corr_ema_eat}
    \begin{minipage}{.95\columnwidth}%
    {\footnotesize {\it Note} 
    The figure report the correlation between ATCs of the eating actions(Breakfast, Lunch, Supper, Snack) and the EMA (weekly average) described in Table~\ref{table:ema_question}. We use all available EMA answers during the period (11 weeks) for each question. On each subfigure, we report the Pearson correlation coefficient ($r$) and p-value for testing non-correlation, and the blue line represents the fitted simple linear regression.
    }
    \end{minipage}%
\end{figure}

\begin{figure}[ht]
     \centering
     \includegraphics[width=.9\linewidth]{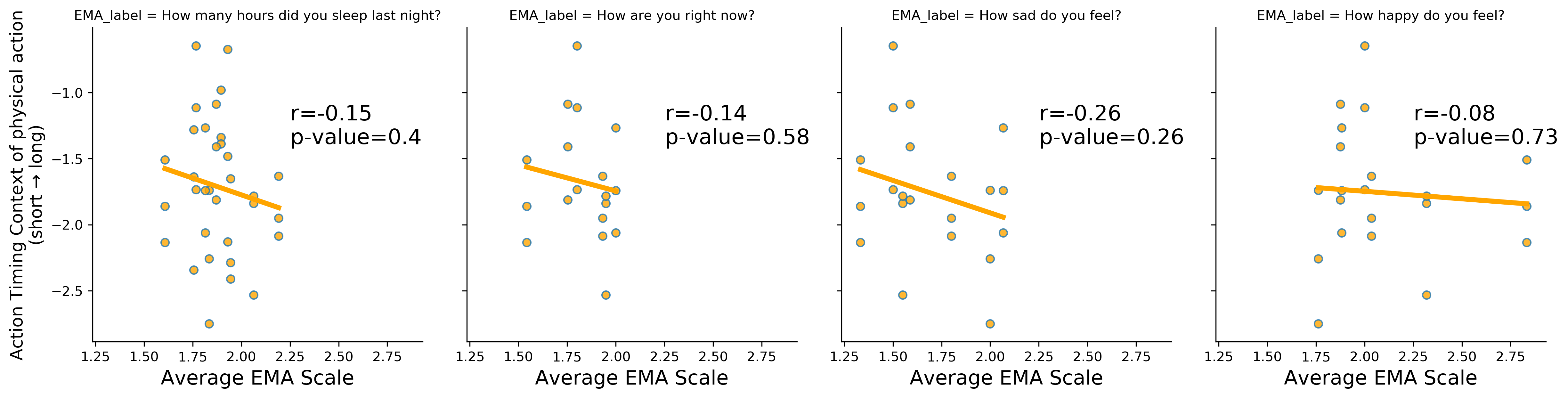}
     \caption{Correlations between ATCs of the physical actions and Physiological state}\label{fig:studentlife_corr_ema_phy}
    \begin{minipage}{.95\columnwidth}%
    {\footnotesize {\it Note}: The figure reports the correlation between ATCs of the physical actions (Walking, Running, Other Activity) and the EMA (weekly average) described in Table~\ref{table:ema_question}.  We use all available EMA answers during the period (11 weeks). On each subfigure, we report the Pearson correlation coefficient ($r$) and p-value for testing non-correlation, and the blue line represents the fitted simple linear regression. 
    }
    \end{minipage}%
\end{figure}

Figure~\ref{fig:studentlife_weekly} shows ATC transitions. Physical behavior ATCs remain stable, whereas eating behavior ATCs fluctuate around Weeks 4–5, coinciding with mid‑terms. This suggests students compress time intervals between eating actions during mid‑terms to study. After Week 6, eating behavior ATCs plateau. For reference, popular apps (Gmail and YouTube) lie between physical and eating behaviors. Also, the figure indicates that external events affect some actions: eating behavior changes with exams, whereas physical behavior does not. Figure~\ref{fig:studentlife_box} corroborates this, showing shorter ATC ranges for physical behavior and longer ranges for eating behavior. Finally, we examine correlations between ATCs and EMA responses. Figure~\ref{fig:studentlife_corr_ema_eat} shows that eating ATCs correlate with mood: they are positively correlated with happiness and negatively with sadness, but not with sleep duration or general mood. Figure~\ref{fig:studentlife_corr_ema_phy} shows no clear correlations between physical ATCs and EMA. Although these are simple correlations, the results suggest that inter‑temporal context may capture aspects of users’ psychological states.

\section{Conclusion \& Discussion}\label{sec:conclusion}

In this study, we proposed a framework for analyzing users' actions by considering the inter-temporal context, based on the idea that time intervals between actions correspond to users' cognitive states. The proposed framework examines the cognitive state from the distribution of time intervals.
To achieve this, we first discretized time intervals based on mixture distribution estimation. Then, we learned embedding vectors of actions using discretized time intervals and action sequences. We proposed ATC, an inter-temporal context index for each action, using the resulting low-dimensional action representation. Using our proposed framework, we conducted empirical studies on user behavior with three different datasets. We found that ATC captures actions, identifies differences in behavior among users, and reveals how the inter-temporal context changes depending on the situation. Analysis showed that ATC provides a unified and interpretable measure of inter-temporal context.

While this study demonstrated that incorporating time intervals into human behavior analysis leads to a unified understanding of dynamic human behavior, it also opens up challenging questions and suggests several further directions. First, we need to analyze the extent to which our framework captures human cognitive changes. The simple analysis with EMA data showed promising results, but we need to conduct a more rigorous analysis on this point. For this direction, we must build a dataset combining behavioral data with subjects' diachronic psychological changes for such analysis. Second, this interdisciplinary study could open up a new research topic. The dual-process theory sees human thinking as switching between two modes, but by using large, complex data and our model, we might better understand how people move between these thinking states, not as a simple switch but as a gradual and continuous shift. Finally, for some tasks, constructing a vector representation of a user based on their actions is necessary, but this paper only focused on each action. It is natural to calculate a user-behavior vector representing user behavior from the obtained action embedding vectors. Just as existing research obtains a text vector by averaging the word embeddings of each word in the text, we can calculate the user's feature vector by averaging the action vectors in the user's action sequence. However, this approach loses the dependency between the time points of the user's actions. Therefore, we need a method that preserves the order of vectors to calculate the user's feature vector.

\end{document}